\documentclass[preprintnumbers,
amsmath,amssymb,floatfix,10pt,prd,onecolumn,
superscriptaddress,longbibliography,nofootinbib]{revtex4-2}
\usepackage{bm}
\usepackage{amsfonts}
\usepackage{latexsym}
\usepackage{amsmath}
\usepackage{textcomp}
\usepackage{float}
\usepackage{booktabs}
\usepackage{dcolumn}
\usepackage{dsfont}
\usepackage{ragged2e}
\usepackage{epsfig}
\usepackage[dvipsnames]{xcolor}
\usepackage{hyperref}
\hypersetup{           
	colorlinks=true,                
	breaklinks=true,                
	urlcolor= OrangeRed,                
	linkcolor= magenta,                
	bookmarksopen=false,
	filecolor=black,
	citecolor=red,
	linkbordercolor=blue}
\usepackage{graphicx}
\usepackage{overpic}
\usepackage{orcidlink}
\usepackage[T1]{fontenc}

\begin{document}
	\title{From Big Bang Nucleosynthesis to Late-Time Acceleration in $f(Q,L_m)$ Gravity}
	
   \author{Rajdeep Mazumdar \orcidlink{0009-0003-7732-875X}}
	\email[Corresponding author: ]{rajdeepmazumdar377@gmail.com}
	\affiliation{%
		Department of Physics, Dibrugarh University, Dibrugarh, Assam, India, 786004}

  \author{Kalyan Malakar\orcidlink{0009-0002-5134-1553}}%
\email{kalyanmalakar349@gmail.com}
\affiliation{Department of Physics, Dibrugarh University, Dibrugarh, Assam, India, 786004}
\affiliation{Department of Physics, Silapathar College, Dhemaji, Assam, India, 787059}
	
	\author{Kalyan Bhuyan\orcidlink{0000-0002-8896-7691}}%
	\email{kalyanbhuyan@dibru.ac.in}
	\affiliation{%
		Department of Physics, Dibrugarh University, Dibrugarh, Assam, India, 786004}%
	\affiliation{Theoretical Physics Division, Centre for Atmospheric Studies, Dibrugarh University, Dibrugarh, Assam, India 786004}

\begin{abstract}
We perform a comprehensive investigation of the early-to-late time cosmic evolution within the framework of $f(Q,L_m)$ gravity, characterized by a non-minimal coupling between non-metricity and matter. The model is further tested against a combined set of observational data, including DESI DR2 BAO, previous BAO measurements, cosmic chronometers (CC), and gravitational-wave (GW) standard sirens, using a Markov Chain Monte Carlo (MCMC) approach. Further by incorporating the Big Bang Nucleosynthesis (BBN) freeze-out constraint, we place stringent limits on the model parameters, ensuring consistency with early-Universe physics.  The resulting constraints exhibit strong agreement with observations, with the model successfully describing the transition from decelerated to accelerated expansion. The evolution of the effective equation-of-state parameter, together with statefinder diagnostics and energy conditions, reveals a quintessence-like nature and confirms the physical viability of the model. Overall, the $f(Q,L_m)$ framework emerges as a viable alternative to $\Lambda$CDM, closely reproducing its predictions while allowing controlled deviations in the expansion history.
\end{abstract}
	
	\maketitle
    \textbf{Keywords:} $f(Q,L_m)$ gravity; Non-metricity; Big-Bang Nucleosynthesis; Late-time acceleration; DESI DR2 BAO; GW.

\section{Introduction}\label{s1}
The discovery that the Universe is undergoing a phase of accelerated expansion is a pivotal advancement in modern cosmology. This finding has been corroborated by various independent observational methods, including precise assessments of the Cosmic Microwave Background Radiation (CMBR) \cite{Spergel2003}, Baryon Acoustic Oscillations (BAO), and extensive surveys of large-scale structure (LSS) \cite{Tegmark2004, Cole2005, Eisenstein2005}. Initial indications of this phenomenon originated from the study of Type Ia Supernovae \cite{Riess1998, Perlmutter1999, Astier2006}. When combined, these results demonstrate that the Universe is not only expanding, but its expansion is accelerating over time. Researchers in diverse studies have postulated the presence of a mysterious component called dark energy (DE), which is described as a kind of energy with enormous negative pressure that accounts for around 70\% of the cosmic energy content, in order to explain this accelerated expansion of the universe. The traditional $\Lambda$CDM model, which is in great agreement with an extensive range of observational evidence spanning many cosmic epochs, is based on the simplest and most frequently accepted candidate for DE, the cosmological constant ($\Lambda$). Despite the empirical robustness of the $\Lambda$CDM model, it possesses conceptual flaws, prominently the "cosmological constant problem," which highlights the disparity between observed and theoretically predicted vacuum energy densities \cite{Weinberg1989, Martin2012}. Additionally, the "fine-tuning" and "cosmic coincidence" problems question the subtle calibration of present-day dark energy density and its precise alignment with the current cosmological era. These theoretical challenges related to cosmic acceleration have led to the creation of alternative frameworks that do not require the addition of a fixed cosmological constant ($\Lambda$). These frameworks include dynamic dark energy models with changing equations of state and modifications to Einstein’s gravitational theory. A thorough review of these approaches is available in the referenced literature \cite{za1}--\cite{za10new3}.\\
Modified gravity theories, which extend or reformulate traditional general relativity, seek to offer a more accurate and thorough understanding of the universe's dynamics on large scales \cite{p13}. These theories are particularly effective in addressing major cosmological challenges, including dark energy, dark matter, and the accelerated expansion of the universe in its later stages, issues that conventional general relativity has not been able to resolve satisfactorily. Modified theories of gravity, derived from generalizing the Einstein-Hilbert action, introduce new components related to geometric features such as curvature invariants, torsion, or non-metricity. These enhancements improve the structural complexity of the theories, thereby boosting their predictive accuracy and offering fresh perspectives on the fundamental dynamics of cosmic evolution. In General Relativity (GR), the fundamental concept of spacetime curvature is expressed through the Riemann tensor. One of the significant modified gravity theories examined is the $f(R)$ gravity theory \cite{p14}, which replaces the Ricci scalar $R$ in the Einstein–Hilbert action with a generic function $f(R)$. Alternative geometric formulations of gravity such as $f(T)$ gravity based on torsion \cite{p16} and $f(Q)$ gravity based on non-metricity \cite{p17,p18}, rather than curvature, have also been seen to attract considerable interest. These frameworks naturally give rise to the $f(T)$ and $f(Q)$ theories \cite{p19,p20,p21}, in which gravitational Lagrangian is generalized with arbitrary functions of $T$ and $Q$. Further extensions incorporating additional couplings and functional dependencies have been shown to yield richer and more versatile theoretical structures. Notable examples include $f(R,T)$ \cite{p22}, $f(R,L_m)$ \cite{p23,p24}, $f(T,\tau)$ \cite{p25}, $f(T,\phi)$ \cite{p26}, $f(T,B)$ \cite{p28}, and $f(Q,T)$ gravity \cite{p29}. Such frameworks enable non-minimal geometry–matter couplings, leading to widely novel gravitational dynamics and observable effects. In particular, $f(Q,L_m)$ gravity \cite{pp45}, where the matter Lagrangian $L_m$ couples directly to the non-metricity scalar $Q$ \cite{pp45new1,pp45new2}, also provides a compelling extension. As, this interaction offers a geometric mechanism for late-time cosmic acceleration without invoking exotic dark energy components such as $\Lambda$. Moreover, the flexibility of $f(Q,L_m)$ may allow a unified description of both early- and late-time cosmological evolution, as supported by several recent studies \cite{pp49}. A detailed formulation of $f(Q,L_m)$ gravity, including its field equations, along with cosmological implications has been presented by Hazarika \textit{et al.}~\cite{pp45}. Observational constraints on late-time acceleration on it were explored by K. Myrzakulov \textit{et al.}~\cite{pp46}, while Y. Myrzakulov \textit{et al.}~\cite{pp47,pp48} investigated the impact of bulk viscosity on it's cosmic dynamics. The phase-space behaviour of Barrow Agegraphic Dark Energy in it was analyzed by Samaddar \& Surendra~\cite{pp33}, demonstrating its capability to unify different evolutionary phases. Further studies \cite{pp45,n1} reinforce the potential of $f(Q,L_m)$ gravity in addressing key theoretical and observational challenges, highlighting its rich structure and cosmological viability.\\
Additionally, the Early-Universe physics also imposes stringent constraints on modified gravity models, complementing late-time observational probes. Based on which recently, Samaddar and Singh~\cite{pp49}  studied the viability of baryogenesis within the $f(Q,L_m)$ framework, which highlighted its possible significance for early-Universe physics. And, just like baryogenesis, the Big Bang Nucleosynthesis (BBN) is also an important early universe probe to evaluate the viability of different gravity models (see Refs. \cite{PK2,PK3,PK4} for latest developments). BBN is notably sensitive to the expansion rate during the radiation-dominated period, imposing strict constraints on any alterations in the Hubble parameter, $H(z)$, during early epochs.~\cite{70,71,PK2}. Despite its theoretical appeal, the cosmological viability of $f(Q,L_m)$ gravity has not yet been systematically established across the full cosmic history, especially in the presence of both early-time constraints through probes specially like BBN and late-time constraints through latest high precisions datasets like DESI DR2 BAO and GW. This motivates a comprehensive investigation of its consistency with the above probes. Hence, we present a novel early-to-late cosmological analysis of the $f(Q,L_m)$ gravity model. Where, the BBN freeze-out constraint is incorporated to test its early-Universe viability, while Markov Chain Monte Carlo (MCMC) analyses using DESI DR2 BAO, P-BAO, cosmic chronometer (CC), and gravitational wave (GW) data constrain and evaluate its late-time behavior. This joint approach tightly bounds the model parameters and assesses its consistency with the full expansion history, including possible deviations from $\Lambda$CDM, thereby highlighting the role of non-metricity–matter couplings in cosmic evolution.\\
This paper is organised as follows: In Sec.~\ref{s2}, we outline the $f(Q,L_m)$ framework and the corresponding background cosmological equations. Sec.~\ref{s3} presents the BBN formalism used to impose early-Universe constraints. In Sec.~\ref{s4}, we introduce the specific $f(Q,L_m)$ model and derive the governing equations for cosmic evolution. Sec.~\ref{s5} describes the observational datasets employed, including DESI DR2 BAO, P-BAO, cosmic chronometers (CC), and gravitational wave (GW) data. The main results, including late-time and BBN constraints and cosmological implications of the model under them are presented in Sec.~\ref{s6}. Finally, Sec.~\ref{s7} summarizes the results and future scope.
\section{$f(Q,L_m)$ Gravity and Modified Friedmann Equations}\label{s2}
In $f(Q,L_m)$ gravity the action is expressed as:
\begin{equation}
S=\int f(Q,L_m)\sqrt{-g}\, d^4x,
\label{e1}
\end{equation}
where $f(Q,L_m)$ characterizes a non-minimal coupling between geometry and matter, and $g$ denotes the determinant of the metric tensor. This formulation extends both $f(Q)$ gravity and General Relativity by introducing direct interactions between the non-metricity sector and the matter Lagrangian at the action level. The non-metricity scalar is defined as~\cite{q}
\begin{equation}
Q=-g^{\mu\nu}\left(L^{\beta}{}_{\alpha\mu}L^{\alpha}{}_{\nu\beta}-L^{\beta}{}_{\alpha\beta}L^{\alpha}{}_{\mu\nu}\right),
\end{equation}
where the disformation tensor is given by
\begin{equation}
L^{\beta}{}_{\alpha\gamma}=\frac{1}{2} g^{\beta\eta}
\left( Q_{\gamma\alpha\eta}+Q_{\alpha\eta\gamma}-Q_{\eta\alpha\gamma} \right),
\end{equation}
and the non-metricity tensor takes the form
\begin{equation}
Q_{\gamma\mu\nu}=-\nabla_{\gamma}g_{\mu\nu}
=-\partial_{\gamma}g_{\mu\nu}
+g_{\nu\sigma}\, \tilde{\Gamma}^{\sigma}{}_{\mu\gamma}
+g_{\sigma\mu}\, \tilde{\Gamma}^{\sigma}{}_{\nu\gamma}.
\end{equation}
Here, $\tilde{\Gamma}^{\sigma}{}_{\mu\nu}$ denotes the symmetric teleparallel connection. The traces of the non-metricity tensor are defined as
\begin{equation}
Q_{\beta}=g^{\mu\nu}Q_{\beta\mu\nu}, 
\qquad 
\tilde{Q}_{\beta}=g^{\mu\nu}Q_{\mu\beta\nu}.
\end{equation}
To simplify the structure of the field equations, we introduce the superpotential tensor (non-metricity conjugate)
\begin{widetext}
\begin{equation}
P^{\beta}{}_{\mu\nu} \equiv \frac{1}{4}
\left[
- Q^{\beta}{}_{\mu\nu}
+ 2 Q^{\beta}{}_{(\mu\nu)}
+ Q_{\beta} g_{\mu\nu}
- \tilde{Q}_{\beta} g_{\mu\nu}
- \delta^{\beta}_{(\mu} Q_{\nu)}
\right],
\end{equation}
\end{widetext}
which plays a role analogous to the contortion tensor in teleparallel gravity. In terms of this tensor, the scalar $Q$ can be recast as~\cite{q}
\begin{widetext}
\begin{equation}
Q = - Q_{\beta\mu\nu} P^{\beta\mu\nu}
= -\frac{1}{4}
\left(
- Q_{\beta\nu\rho} Q^{\beta\nu\rho}
+ 2 Q_{\beta\nu\rho} Q^{\rho\beta\nu}
- 2 Q_{\rho} \tilde{Q}^{\rho}
+ Q_{\rho} Q^{\rho}
\right).
\end{equation}
\end{widetext}
Variation of the action with respect to the metric yields the modified field equations
\begin{widetext}
\begin{equation}
2\sqrt{-g}\,\nabla_{\alpha}\!\left(f_{Q}\sqrt{-g}\,P^{\alpha}{}_{\mu\nu}\right)
+ f_{Q}\left(P_{\mu\alpha\beta} Q^{\alpha\beta}{}_{\nu}
- 2 Q_{\alpha\beta\mu} P^{\alpha\beta}{}_{\nu}\right)
+ \frac{1}{2} f\, g_{\mu\nu}
= \frac{1}{2} f_{L_m} \left(g_{\mu\nu} L_{m} - T_{\mu\nu}\right),
\label{e8}
\end{equation}
\end{widetext}
where $f_Q=\partial f/\partial Q$ and $f_{L_m}=\partial f/\partial L_m$. The standard $f(Q)$ limit is recovered for $f(Q,L_m)=f(Q)+2L_m$~\cite{q}. The energy-momentum tensor is defined as
\begin{equation}
T_{\mu\nu}
= g_{\mu\nu}L_m - 2\frac{\partial L_m}{\partial g^{\mu\nu}}.
\end{equation}
Variation with respect to the connection produces an additional dynamical equation
\begin{equation}
\nabla_{\mu}\nabla_{\nu}
\left(4\sqrt{-g}\, f_Q P^{\mu\nu}{}_{\alpha}
+ H^{\mu\nu}{}_{\alpha}\right)=0,
\end{equation}
where the hypermomentum tensor is
\begin{equation}
H^{\mu\nu}{}_{\alpha}
= \sqrt{-g}\, f_{L_m} \frac{\partial L_m}{\partial Y^{\alpha}{}_{\mu\nu}}.
\end{equation}
A defining feature of $f(Q,L_m)$ gravity is the non-conservation of the energy–momentum tensor induced by the non-minimal coupling. Taking the covariant divergence of the field equations leads to
\begin{equation}
\nabla_{\mu} T^{\mu}{}_{\nu} = B_\nu \neq 0,
\end{equation}
which signifies an exchange of energy and momentum between the matter and geometric sectors. Standard conservation is recovered only in the limit $B_\nu=0$. To study cosmological dynamics, we adopt a spatially flat Friedmann–Lemaître–Robertson–Walker (FLRW) background
\begin{equation}
ds^{2} = -dt^{2} + a^{2}(t)\left(dx^{2} + dy^{2} + dz^{2}\right),
\label{metric}
\end{equation}
with Hubble parameter $H=\dot{a}/a$. For this metric, the non-metricity scalar reduces to
\begin{equation}
Q = 6H^{2}.
\end{equation}
Assuming a perfect fluid description, with $\rho$ and $p$ as the energy and pressure of the cosmic fluid, we get:
\begin{equation}
T_{\mu\nu} = (\rho + p)u_{\mu}u_{\nu} + p\, g_{\mu\nu},
\end{equation}
the modified Friedmann equations become~\cite{pp45,pp46,pp47,pp48}
\begin{equation}
3H^{2} = \frac{1}{4 f_Q}\left[f - f_{L_m}(\rho + L_m)\right], 
\label{fried1}
\end{equation}
\begin{equation}
\dot{H} + 3H^{2} + \frac{\dot{f}_Q}{f_Q}H = \frac{1}{4 f_Q}\left[f + f_{L_m}(p - L_m)\right].
\label{fried2}
\end{equation}
where, we take $8\pi G = 1$. Now, Eq. (\ref{fried1}) in GR equivalent form will be:
\begin{equation}
    3H^2 = \rho + \rho_{\text{DE}},
\end{equation}
where $\rho_{\text{DE}}$ is the dark energy density, which can be defined as:
\begin{equation}
    \rho_{DE} = \frac{1}{4 f_Q}\left[f - f_{L_m}(\rho + L_m)\right] -\rho,
    \label{de}
\end{equation}

\section{Big Bang Nucleosynthesis Constraints} \label{s3}
In this section, we outline the connection between the fractional variation of the freeze-out temperature and Big Bang Nucleosynthesis (BBN) constraints in a concise way. During the radiation-dominated epoch, the first Friedmann equation in standard General Relativity can be approximated as \cite{70,71}:
\begin{equation}
H^{2} \simeq \frac{\rho_{r}}{3M_{p}^{2}} \equiv H_{\rm GR}^{2},
\end{equation}
where $\rho_r$ denotes the radiation energy density and $M_p=(8\pi G)^{-1/2}$ is the reduced Planck mass. The radiation density is given by
\begin{equation}
\rho_{r} = \frac{\pi^{2}}{30} g_{} T^{4},
\end{equation}
with $g_\simeq 10$ during the BBN epoch. Substituting, the Hubble parameter becomes
\begin{equation}
H(T) = \left(\frac{4\pi^{3}g_{*}}{45}\right)^{1/2}\frac{T^{2}}{M_{p}}.
\end{equation}
Radiation conservation implies $a(t)\propto t^{1/2}$, leading to $H=1/(2t)$ and the relation
\begin{equation}
\frac{1}{t} = \left(\frac{16\pi^{3}g_{*}}{45}\right)^{1/2}\frac{T^{2}}{M_{p}},
\end{equation}
or equivalently $T\propto t^{-1/2}$. During BBN, neutron–proton conversion processes govern particle interactions \cite{70,71}, with the total interaction rate expressed as
\begin{equation}
\Gamma_{\rm tot}(T) = 8\left(12T^{2}+6QT+Q^{2}\right)AT^{3},
\end{equation}
where $Q=m_n-m_p$ and $A=1.02\times10^{-11},{\rm GeV}^{-4}$ \cite{72,73}. The primordial helium mass fraction is:
\begin{equation}
Y_{p}=\frac{2\lambda,x(T_{f})}{1+x(T_{f})},
\end{equation}
with $x(T_f)=\exp(-Q/T_f)$ and $\lambda=\exp[(T_f-T_n)/\tau]$. The freeze-out temperature $T_f$ is determined from the condition
\begin{equation}
H(T_f)=\Gamma_{\rm tot}(T_f)\simeq c_q T_f^5,
\end{equation}
where $c_q=96A$ \cite{bbn54,72,73}. This yields
\begin{equation}
T_{f}=\left(\frac{4\pi^{3}g_{*}}{45,c_{q}^{2}M_{p}^{2}}\right)^{1/6}.
\end{equation}
In modified cosmology, deviations in $H$ induce a shift $T_f \rightarrow T_f + \Delta T_f$, leading to a correction in helium abundance:
\begin{equation}
\Delta Y_{p}
=Y_{p}\left[\frac{1-Y_{p}}{2\lambda}\ln\left(\frac{2\lambda}{Y_{p}}-1\right)
-\frac{2T_{f}}{\tau}\right]\frac{\Delta T_{f}}{T_{f}}.
\end{equation}
Observationally, $Y_p=0.2476$ with $|\Delta Y_p|<10^{-4}$ \cite{76}. In extended gravity models, the Friedmann equation becomes:
\begin{equation}
3M_{p}^{2}H^{2}=\rho_{m}+\rho_{r}+\rho_{\rm DE}.
\end{equation}
During BBN, $\rho_m$ is negligible, giving
\begin{equation}
H = H_{\rm GR}\left(1+\frac{\rho_{\rm DE}}{\rho_{r}}\right)^{1/2}.
\end{equation}
For $\rho_{\rm DE}\ll\rho_r$,
\begin{equation}
H \simeq H_{\rm GR}\left(1+\frac{1}{2}\frac{\rho_{\rm DE}}{\rho_{r}}\right).
\end{equation}
This modification leads to:
\begin{equation}
\frac{\Delta T_{f}}{T_{f}} \simeq
\frac{\rho_{\rm DE}}{\rho_{r}}
\frac{H_{\rm GR}}{10,c_{q}T_{f}^{5}},
\label{deltaTf}
\end{equation}
subject to the observational bound
\begin{equation}
\left|\frac{\Delta T_{f}}{T_{f}}\right| < 4.7\times10^{-4}.
\label{Tf_bound}
\end{equation}
This constraint, derived from helium abundance measurements, provides a stringent test of early-Universe physics. Thus, BBN serves as a powerful probe to constrain modified gravity models by linking theoretical predictions with observational data.
\section{Cosmological Model}\label{s4}
Within the framework of $f(Q,L_m)$ gravity, we explore an extended gravity model in this study that is specified as:
\begin{equation}
f(Q,\mathcal{L}_m) = \alpha Q + \beta \mathcal{L}_m^{\,n} + \lambda ,
\label{m}
\end{equation}
where the independent model parameters are $\alpha$, $\beta$, $\lambda$, and $n$. For which we take the matter Lagrangian as $\mathcal{L}_m = \rho$. This power-law form introduces a non-minimal coupling between the matter Lagrangian and non-metricity, extending the minimal scenario and enabling richer cosmological dynamics. The parameter $n$ controls the strength of this coupling.
It is worth noting that the $f(Q,\mathcal{L}_m)$ model used in this study shares some comparable similarities with those investigated by K. Myrzakulov et al. \cite{pp46} and Y. Myrzakulov et al. \cite{pp47, n1}. However, our present formulation extends those previous studies in some key aspects. First, we parametrized the coupling strength of the non-metricity scalar $Q$ by a free parameter $\alpha$, hence allowing controlled deviations from the canonical STEGR normalization. Second, the addition of a constant term $\lambda$ in the function may effectively act as a dynamical vacuum contribution, which can modulate or influence the late-time cosmic acceleration. When combined with new high-precision observational datasets, these modifications may offer more theoretical flexibility, resulting in richer cosmic dynamics and enabling a broader spectrum of expansion histories beyond those examined in previous $f(Q,\mathcal{L}_m)$ frameworks.\\
Now, Eq. (\ref{de}) for the above model gives us:
\begin{equation}
\rho_{DE} = \frac{2^{-n-1} 3^{1-n} \left(H^2\right)^{1-n} \left(6^n \left(H^2\right)^n+6 \lambda -\beta  (2 n-1) \rho ^n\right)}{n}-\rho
\label{e}
\end{equation}
During the BBN era we can take $H \sim H_{GR}$ and $\rho \sim \rho_{r}$, using which in Eqs. (\ref{e}) and (\ref{deltaTf}) along with the aforementioned expressions of $H_{GR}$ and $\rho_{r}$, we obtain the fractional change in freezing out temperature as:
\begin{equation}
\frac{\Delta T_{f}}{T_{f}}=\frac{\left(\frac{3}{10 \pi }\right)^{2/3} c_q T_f^2 M_p \sqrt[6]{\frac{g_*}{c_q^2 M_p^2}} \left(\frac{2^{-n} \pi ^{-2 n} {\left(\frac{g_* T_f^4}{M_p^2}\right)}^{-n} \left(\pi ^{2 n} \left(2^n {\left(\frac{g_* T_f^4}{M_p^2}\right)}^n-\tilde{\beta}  (2 n-1) {\left(g_* T_f^4\right)}^n\right)+\tilde{\lambda}  5^n 6^{n+1}\right)}{{M_p}^2 n}-2\right)}{2 \sqrt{g_*}}
\label{eqtf}
\end{equation}
where, $\tilde{\lambda} = \frac{\lambda}{6 \alpha}$ and $\tilde{\beta} = \frac{\beta}{\alpha}$. Eq. (\ref{eqtf}) will be used to analysis the early universe through the validity of BBN in the given model.\\
For the given model using Eqs. (\ref{fried1}) and (\ref{fried2}) the Hubble parameter in terms of the redshift $z$ can be obtained as:
\begin{equation}
H^2(z) =
H_0^2 (1+z)^{\frac{3n(1+\omega)}{2n-1}}
+ \tilde{\lambda} \left(1-(1+z)^{\frac{3n(1+\omega)}{2n-1}}\right)
,
\label{Hz}
\end{equation}
where, $\tilde{\lambda} = \frac{\lambda}{6\alpha}$ having dimension equivalent to $H^2$. Eq.~(\ref{Hz}) will be used to explore the late-time dynamics with constraints from observational datasets like DESI DR2 BAO and GW.
\section{Observational Data and Methodology}\label{s5}
We constrain the free parameters of the proposed cosmological model using a combination of cosmic chronometer (CC), Baryon Acoustic Oscillation (BAO), and gravitational-wave (GW) standard siren data. The BAO measurements consist of the latest high-precision DESI DR2 dataset~\cite{ref90} as well as an earlier compilation (P-BAO) from surveys such as SDSS, WiggleZ, and DES~\cite{d113}--\cite{d79}. Here, CC data provide direct estimates of the Hubble parameter $H(z)$ through differential age measurements of passively evolving galaxies~\cite{x38}. In addition, GW observations serve as an independent probe via standard sirens, using publicly available data from the LIGO–Virgo–KAGRA catalogues (GWTC-1, GWTC-2, GWTC-2.1, and GWTC-3)~\cite{g1,g2}. Together, these datasets enable a robust reconstruction of the cosmic expansion history across a wide redshift range. To evaluate the model, we perform three separate analyses:
\begin{itemize}
\item DESI,
\item P-BAO,
\item DESI + P-BAO + CC + GW.
\end{itemize}
For the BAO analysis, we employ the observables such as Hubble parameter \(H(z)\), the Hubble distance \(D_H(z)\), the transverse comoving distance \(D_M(z)\), and the volume-averaged distance \(D_V(z)\). These quantities are normalized with respect to the sound horizon at the drag epoch \(r_d\), which is given as:
\begin{equation}
r_d = \int_{z_d}^{\infty} \frac{c_s(z')}{H(z')} \, dz',
\end{equation}
where \(c_s(z)\) denotes the sound speed of the baryon--photon fluid. The relevant distance measures are given by
\begin{equation}
D_H(z) = \frac{c}{H(z)},
\end{equation}
\begin{equation}
D_M(z) = c \int_0^z \frac{dz'}{H(z')},
\end{equation}
and
\begin{equation}
D_V(z) = \left[z\, D_M^2(z)\, D_H(z)\right]^{1/3}.
\end{equation}
For convenience, we limit our study to the previously described datasets, even though additional probes, such Type Ia supernovae, can potentially offer useful restrictions.\\
For the GW analysis, we employ luminosity distance measurements derived from the GWTC catalogs, adopting median values along with their corresponding \(1\sigma\) uncertainties obtained through Bayesian inference. The redshift information is either obtained from electromagnetic counterparts or statistically inferred for binary black hole mergers. In this framework, GW observations act as distance--redshift probes, with the theoretical luminosity distance given by
\begin{equation}
d_L(z) = (1+z)\int_0^z \frac{dz'}{H(z')}.
\end{equation}
The corresponding distance modulus is defined as
\begin{equation}
\mu = 5 \log_{10} d_L(z) + \mu_0,
\end{equation}
where
\begin{equation}
\mu_0 = 5 \log_{10} \left( \frac{1}{H_0\, \mathrm{Mpc}} \right) + 25,
\end{equation}
and \(H_0\) denotes the present Hubble parameter. The corresponding observational data points along with their uncertainties are listed in Tables~\ref{tab1a}--\ref{tab1c}. Finally, the best-fit model parameters are obtained through standard \(\chi^2\) minimization. The corresponding \(\chi^2\) functions employed for each dataset are defined as follows:
\begin{table}[h!]
\centering
\begin{tabular}{|ccc|c||ccc|c|}
\hline
 \multicolumn{4}{|c||}{\textbf{P-BAO}} & \multicolumn{4}{c|}{\textbf{CC}} \\
\hline
$z$ & $H(z)$ & $\sigma_H$ & Ref & $z$ & $H(z)$ & $\sigma_H$ & Ref \\
\hline
0.24 & 79.69 & 2.99 & \cite{d113} & 0.07 & 69.00 & 19.60 & \cite{cc54} \\
 0.30 & 81.70 & 6.22 & \cite{d114} & 0.09 & 69.00 & 12.00 & \cite{cc55} \\
 0.31 & 78.17 & 6.74 & \cite{d115} & 0.12 & 68.60 & 26.20 & \cite{cc54} \\
0.34 & 83.17 & 6.74 & \cite{d113} & 0.17 & 83.00 & 8.00 & \cite{cc55} \\
 0.35 & 82.70 & 8.40 & \cite{d116} & 0.179 & 75.00 & 4.00 & \cite{cc56} \\
0.36 & 79.93 & 3.39 & \cite{d115} & 0.199 & 75.00 & 5.00 & \cite{cc56} \\
0.38 & 81.50 & 1.90 & \cite{d5} & 0.20 & 72.90 & 29.60 & \cite{cc54} \\
 0.40 & 82.04 & 2.03 & \cite{d115} & 0.27 & 77.00 & 14.00 & \cite{cc55} \\
 0.43 & 86.45 & 3.68 & \cite{d113} & 0.28 & 88.80 & 36.60 & \cite{cc54} \\
 0.44 & 82.60 & 7.80 & \cite{d74} & 0.352 & 83.00 & 14.00 & \cite{cc56} \\
0.44 & 84.81 & 1.83 & \cite{d115} & 0.3802 & 83.00 & 13.50 & \cite{cc58} \\
0.48 & 87.79 & 2.03 & \cite{d115} & 0.4 & 95.00 & 17.00 & \cite{cc55} \\
0.56 & 93.33 & 2.32 & \cite{d115} & 0.4004 & 77.00 & 10.20 & \cite{cc58} \\
0.57 & 87.60 & 7.80 & \cite{d10} & 0.4247 & 87.10 & 11.20 & \cite{cc58} \\
 0.57 & 96.80 & 3.40 & \cite{d117} & 0.4497 & 92.80 & 12.90 & \cite{cc58} \\
0.59 & 98.48 & 3.19 & \cite{d115} & 0.47 & 89.00 & 50.00 & \cite{cc59} \\
0.60 & 87.90 & 6.10 & \cite{d74} & 0.4783 & 80.90 & 9.00 & \cite{cc58} \\
 0.61 & 97.30 & 2.10 & \cite{d5} & 0.48 & 97.00 & 62.00 & \cite{cc59} \\
 0.64 & 98.82 & 2.99 & \cite{d115} & 0.593 & 104.00 & 13.00 & \cite{cc56} \\
0.978 & 113.72 & 14.63 & \cite{d118} & 0.68 & 92.00 & 8.00 & \cite{cc56} \\
1.23 & 131.44 & 12.42 & \cite{d118} & 0.781 & 105.00 & 12.00 & \cite{cc56} \\
 1.48 & 153.81 & 6.39 & \cite{d79} & 0.875 & 125.00 & 17.00 & \cite{cc56} \\
1.526 & 148.11 & 12.71 & \cite{d118} & 0.88 & 90.00 & 40.00 & \cite{cc59} \\
 1.944 & 172.63 & 14.79 & \cite{d118} & 0.9 & 117.00 & 23.00 & \cite{cc55} \\
 2.30 & 224.00 & 8.00 & \cite{d119} & 1.037 & 154.00 & 20.00 & \cite{cc56} \\
2.36 & 226.00 & 8.00 & \cite{d120} & 1.3 & 168.00 & 17.00 & \cite{cc55} \\
2.40 & 227.80 & 5.61 & \cite{d121} & 1.363 & 160.00 & 33.60 & \cite{cc60} \\
 &  &  &  & 1.43 & 177.00 & 18.00 & \cite{cc55} \\
 &  &  &  & 1.53 & 140.00 & 14.00 & \cite{cc55} \\
 &  &  &  & 1.75 & 202.00 & 40.00 & \cite{cc55} \\
 &  &  &  & 1.965 & 186.50 & 50.00 & \cite{cc60} \\
\hline
\end{tabular}
\caption{Summarised $H(z)$ and its uncertainty at redshift $z$ from the Cosmic Chronometer (CC) and P-BAO datasets, measured in units of $km s^{-1} Mpc^{-1}$.}
\label{tab1a}
\end{table}

\begin{table}[h]
\centering
\renewcommand{\arraystretch}{1.4}
\begin{tabular}{lcccc}
\hline\hline
Tracer & $z_{\mathrm{eff}}$ & $D_M/r_d$ & $D_H/r_d$ & $D_V/r_d$ \\
\hline
BGS        & 0.295 & --                  & --                  & $7.942 \pm 0.075$ \\
LRG1       & 0.510 & $13.588 \pm 0.167$  & $21.863 \pm 0.425$  & $12.720 \pm 0.099$ \\
LRG2       & 0.706 & $17.351 \pm 0.177$  & $19.455 \pm 0.330$  & $16.050 \pm 0.110$ \\
LRG3+ELG1  & 0.934 & $21.576 \pm 0.152$  & $17.641 \pm 0.193$  & $19.721 \pm 0.091$ \\
ELG2       & 1.321 & $27.601 \pm 0.318$  & $14.176 \pm 0.221$  & $24.252 \pm 0.174$ \\
QSO        & 1.484 & $30.512 \pm 0.760$  & $12.817 \pm 0.516$  & $26.055 \pm 0.398$ \\
Ly-$\alpha$ QSO & 2.330 & $38.988 \pm 0.531$ & $8.632 \pm 0.101$  & $31.267 \pm 0.250$ \\
\hline\hline
\end{tabular}
\caption{Summary of DESI BAO DR2 measurements at different effective redshifts employed for the study \cite{ref90}.}
\label{tab1b}
\end{table}

\begin{table}[h!]
\centering
\renewcommand{\arraystretch}{1.5}
\begin{tabular}{cccc}
\hline\hline
$z_{\rm eff}$ & Observable & Value & Reference \\
\hline

0.38  & $D_M/r_d$ & $10.272 \pm 0.135 \pm 0.074$ & \cite{w59} \\
0.51  & $D_M/r_d$ & $13.378 \pm 0.156 \pm 0.095$ & \cite{w59} \\
0.61  & $D_M/r_d$ & $15.449 \pm 0.189 \pm 0.108$ & \cite{w59} \\
0.698 & $D_M/r_d$ & $17.65 \pm 0.30$ & \cite{w56} \\
1.48  & $D_M/r_d$ & $30.21 \pm 0.79$ & \cite{w57} \\
2.30  & $D_M/r_d$ & $37.77 \pm 2.13$ & \cite{w55} \\
2.40  & $D_M/r_d$ & $36.60 \pm 1.20$ & \cite{w60} \\

\hline

0.698 & $D_H/r_d$ & $19.77 \pm 0.47$ & \cite{w56} \\
1.48  & $D_H/r_d$ & $13.23 \pm 0.47$ & \cite{w57} \\
2.30  & $D_H/r_d$ & $9.07 \pm 0.31$ & \cite{w55} \\
2.40  & $D_H/r_d$ & $8.94 \pm 0.22$ & \cite{w60} \\

\hline

0.15 & $D_V/r_d$ & $4.473 \pm 0.159$ & \cite{w54} \\
0.44 & $D_V/r_d$ & $11.548 \pm 0.559$ & \cite{w62}  \\
0.60 & $D_V/r_d$ & $14.946 \pm 0.680$ & \cite{w62}  \\
0.73 & $D_V/r_d$ & $16.931 \pm 0.579$ & \cite{w62}  \\
1.52 & $D_V/r_d$ & $26.005 \pm 0.995$ & \cite{w58}  \\

\hline\hline
\end{tabular}

\caption{Summary of previous BAO (P-BAO) measurements at different effective redshifts used in this study.}
\label{tab1c}
\end{table}
\subsection{CC Dataset}
The cosmic chronometer (CC) dataset gives direct estimates of the Hubble parameter $H(z)$ with independent Gaussian errors. For which the corresponding chi-square statistic is given by:
\begin{equation}
\chi^2_{\rm CC}(\theta) = \sum_{j} \frac{\left[H^{\rm obs}(z_j) - H^{\rm th}(z_j,\theta)\right]^2}{\sigma_{H,j}^2},
\end{equation}
where $H^{\rm obs}(z_j)$ and $\sigma_{H,j}$ denote the observed Hubble parameter and its associated uncertainty at redshift $z_j$, while $H^{\rm th}(z_j,\theta)$ represents the theoretical prediction for a given parameter set $\theta$.
\subsection{DESI Dataset}
Due to the presence of correlations among BAO observables in the DESI DR2 BAO dataset, a full covariance matrix approach is required. The corresponding chi-square function is given by:
\begin{equation}
\chi^2_{\rm DESI}(\theta) =
\sum_{i,j}
\left(X^{\rm obs}_i - X^{\rm th}_i(\theta)\right)
C^{-1}_{ij}
\left(X^{\rm obs}_j - X^{\rm th}_j(\theta)\right),
\label{chi_desi}
\end{equation}
where $X^{\rm obs} = \{D_M(z)/r_d,\, D_H(z)/r_d,\, D_V(z)/r_d\}$ represents the observed BAO quantities, $X^{\rm th}$ denotes their theoretical counterparts, and $C^{-1}_{ij}$ is the inverse covariance matrix supplied by the DESI collaboration.

\subsection{P-BAO Dataset}
Unlike the aforementioned DESI DR2 dataset, the previous BAO (P-BAO) compilation used here combines measurements from multiple surveys. Data points at different effective redshifts are treated as independent; however, quantities such as $D_M/r_d$ and $D_H/r_d$ evaluated at the same redshift are typically correlated. In such cases, the covariance matrices reported in the original analyses are included in the likelihood construction~\cite{w56,w57,w55,w60}. The total chi-square for the P-BAO dataset is therefore expressed as:
\begin{widetext}
\begin{equation}
\chi^2_{\rm P\mbox{-}BAO}(\theta) =
\sum_{i}
\Delta \mathbf{D}_i^{T} \, \mathbf{C}_i^{-1} \, \Delta \mathbf{D}_i
+ \sum_{j}
\frac{\left[X^{\rm obs}_j - X^{\rm th}_j(\theta)\right]^2}{\sigma_{X,j}^2},
\end{equation}
\end{widetext}
where the data vector for correlated measurements is defined as
\begin{equation}
\Delta \mathbf{D}_i =
\begin{bmatrix}
\dfrac{D_M^{\rm obs}(z_i)}{r_d} - \dfrac{D_M^{\rm th}(z_i,\theta)}{r_d} \\
\dfrac{D_H^{\rm obs}(z_i)}{r_d} - \dfrac{D_H^{\rm th}(z_i,\theta)}{r_d}
\end{bmatrix},
\end{equation}
and \(\mathbf{C}_i\) denotes the covariance matrix associated with the correlated pair \(\left(D_M/r_d,\, D_H/r_d\right)\) at redshift \(z_i\). For the uncorrelated data points, we consider $ X^{\rm obs}_j = \left\{ \frac{D_M(z_j)}{r_d},\, \frac{D_H(z_j)}{r_d},\, \frac{D_V(z_j)}{r_d},\, H(z_j) \right\}$,
where \(X^{\rm th}_j(\theta)\) represents the corresponding theoretical predictions, and $
\sigma_{X,j} = \left\{ \sigma_{D_M/r_d,j},\, \sigma_{D_H/r_d,j},\, \sigma_{D_V/r_d,j},\, \sigma_{H,j} \right\}$ denotes the associated observational uncertainties.

\subsection{GW Dataset}
Here, the gravitational-wave (GW) dataset is incorporated through luminosity distance measurements, expressed via the distance modulus. For which corresponding chi-square function is defined as:
\begin{equation}
\chi^2_{\rm GW}(\theta) = \sum_{i=1}^{N_{\rm GW}} 
\frac{\left[\mu^{\rm obs}(z_i) - \mu^{\rm th}(z_i;\theta)\right]^2}
{\sigma_{\mu,i}^2},
\end{equation}
where $\mu^{\rm obs}(z_i)$ and $\sigma_{\mu,i}$ represent the observed distance modulus and its associated uncertainty for each GW event, while $\mu^{\rm th}(z_i;\theta)$ denotes the theoretical prediction for a given parameter set $\theta$.\\
Based of above definitions the total chi-square for each analysis is given by:
\begin{equation}
\chi^2_{\rm tot} =
\begin{cases}
\chi^2_{\rm DESI}, & \text{DESI}, \\[6pt]
\chi^2_{\rm P\mbox{-}BAO}, & \text{P-BAO}, \\[6pt]
\chi^2_{\rm DESI} + \chi^2_{\rm P\mbox{-}BAO} + \chi^2_{\rm GW} + \chi^2_{\rm CC}, & \text{DESI + P-BAO + GW + CC}.
\end{cases}
\end{equation}
Based on the \(\chi^2\) functions and datasets described above, a Markov Chain Monte Carlo (MCMC) method is employed to explore the parameter space and obtain the posterior distributions of the model parameters $n$, $\tilde{\gamma}$ and \(H_0\). We adopt uniform priors given by \(H_0 \in [60, 80]\), $\tilde{\gamma} \in [3000,4000]$ and \(n \in [0, 2]\). The MCMC chains are initialized at \((H_0, \Omega_{m0}) = (70,1,3400)\). To assess the goodness of fit and compare the proposed model with the standard $\Lambda$CDM cosmology, we employ crucial statistical estimators, including the minimum chi-square ($\chi^2_{\min}$), the Akaike Information Criterion (AIC), and the Bayesian Information Criterion (BIC). Where, lower values of such estimators point at the statistical superiority of a given model as compare to other models. One can consult Ref. \cite{R1} and Ref. \cite{R2} for more detailed discussion on their applications.
\section{Constraints and Cosmological implications}\label{s6} 
\begin{figure*}[t]
\centerline{
\includegraphics[width=0.38\textwidth]{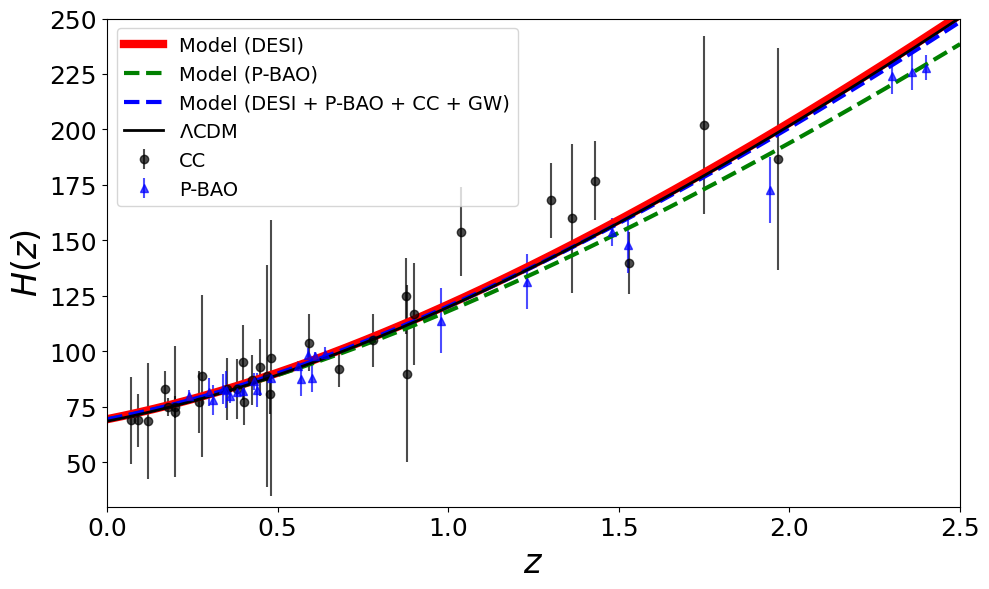}
\includegraphics[width=0.38\textwidth]{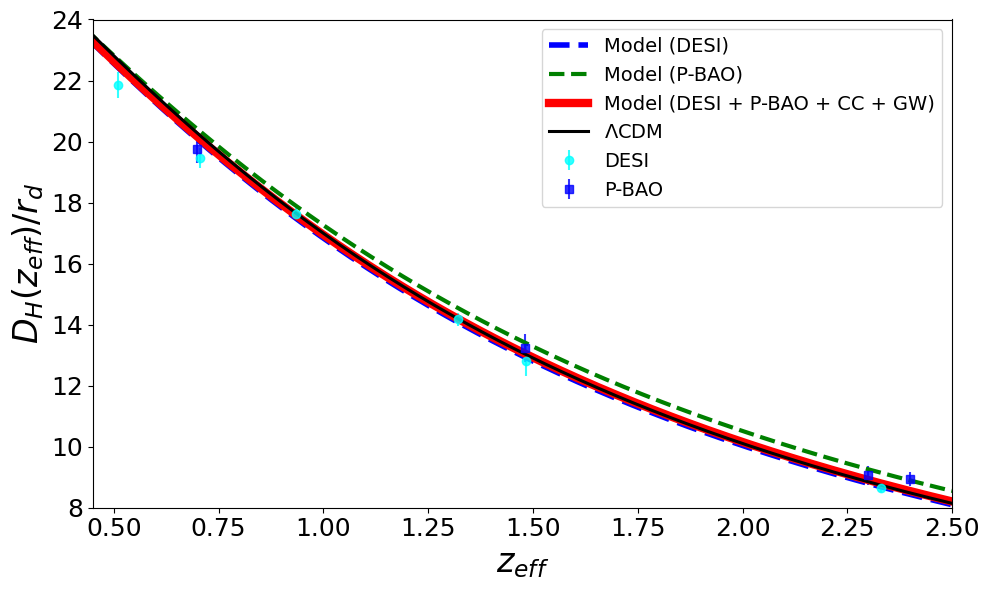}}
\centerline{
\includegraphics[width=0.38\textwidth]{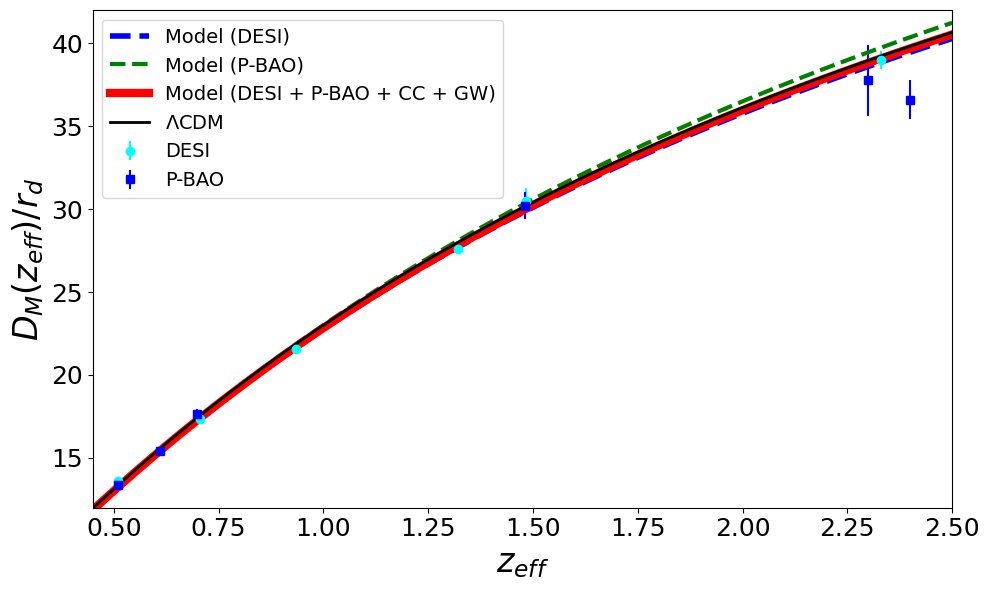}
\includegraphics[width=0.38\textwidth]{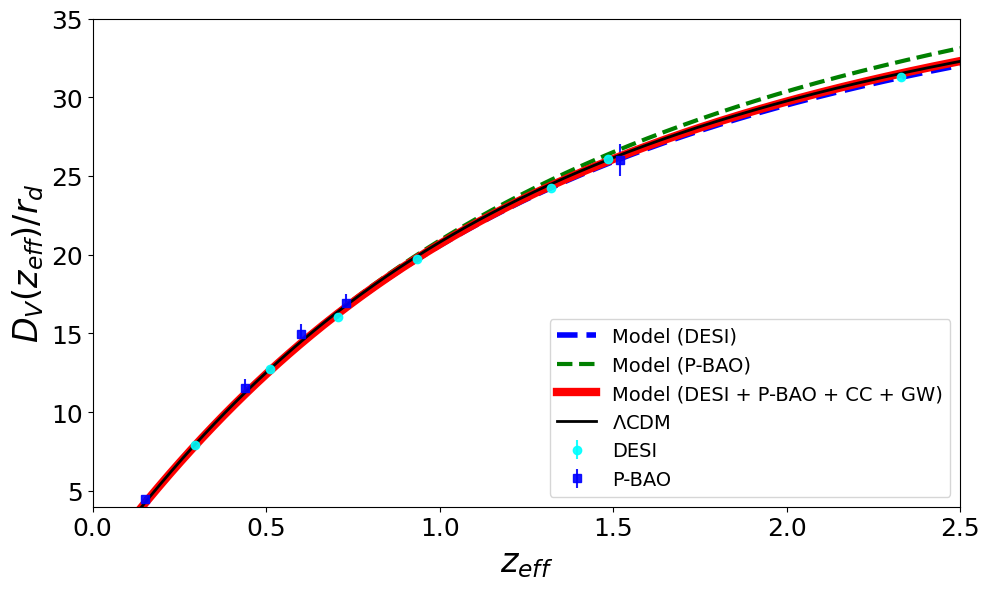}}
\centerline{
\includegraphics[width=0.38\textwidth]{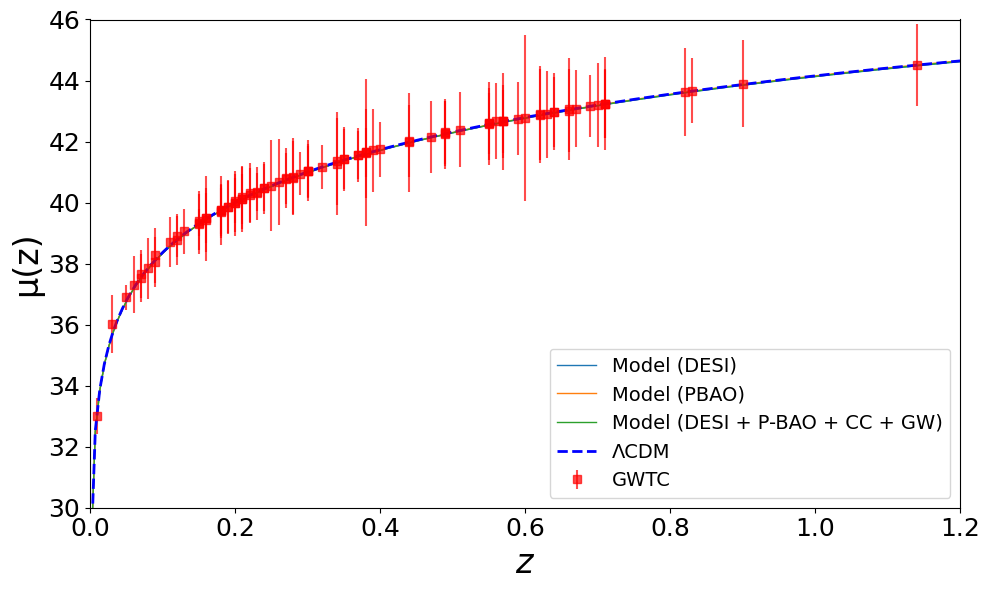}}
\caption{For the model that best matches the observational data, the evolution of $H(z)$ and $\mu$ as functions of $z$, as well as $D_M(z_{\mathrm{eff}})/r_d$, $D_H(z_{\mathrm{eff}})/r_d$, and $D_V(z_{\mathrm{eff}})/r_d$ as functions of $z_{\mathrm{eff}}$, are shown above. Every plot has a comparison with the $\Lambda$CDM model for completeness.}
\label{f1}
\end{figure*}

\begin{figure*}[h!]
\centerline{
\includegraphics[width=0.4\textwidth]{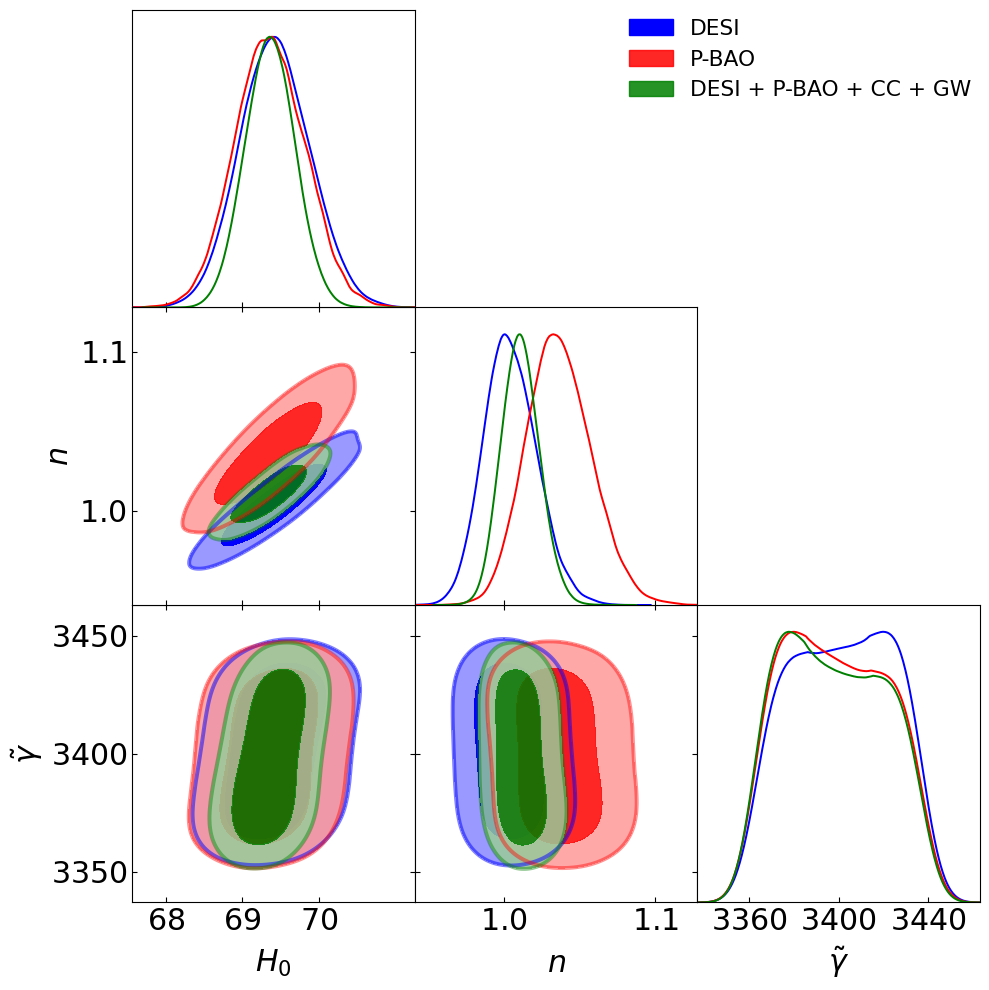}}
\caption{2-d contour subplots for the parameters $H_0$, $n$, and $\tilde{\gamma}$ showing the 1-$\sigma$ and 2-$\sigma$ confidence regions (corresponding to 68\% and 95\% confidence levels).}
\label{f2}
\end{figure*}

\begin{figure*}[htb]
\centerline{
\includegraphics[width=0.8\textwidth]{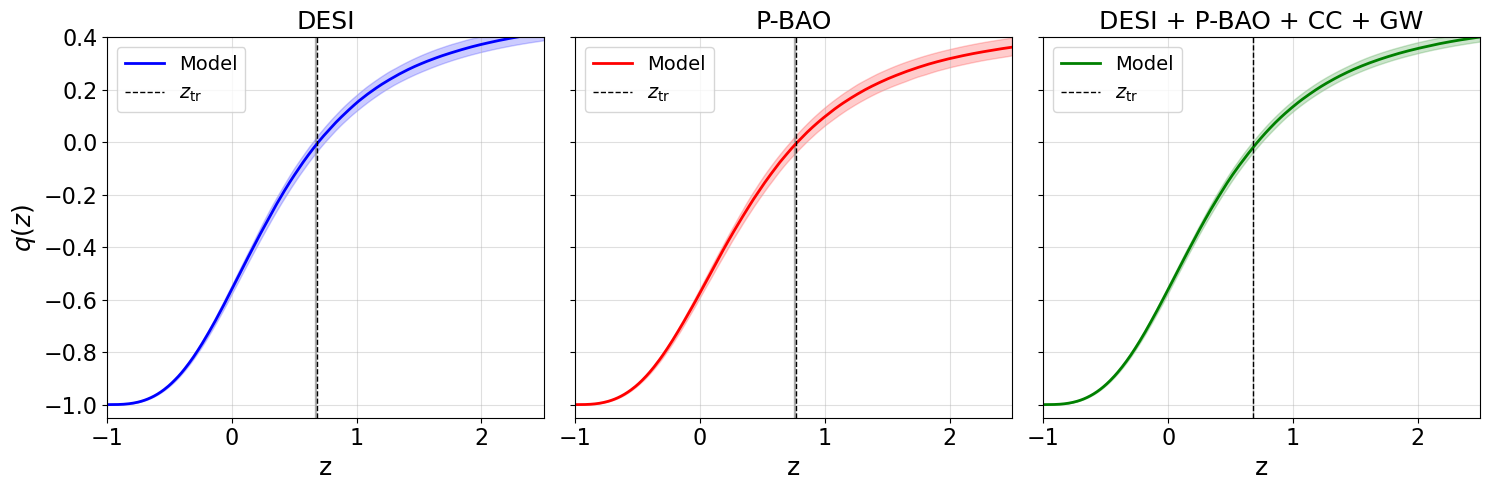}}
\centerline{
\includegraphics[width=0.8\textwidth]{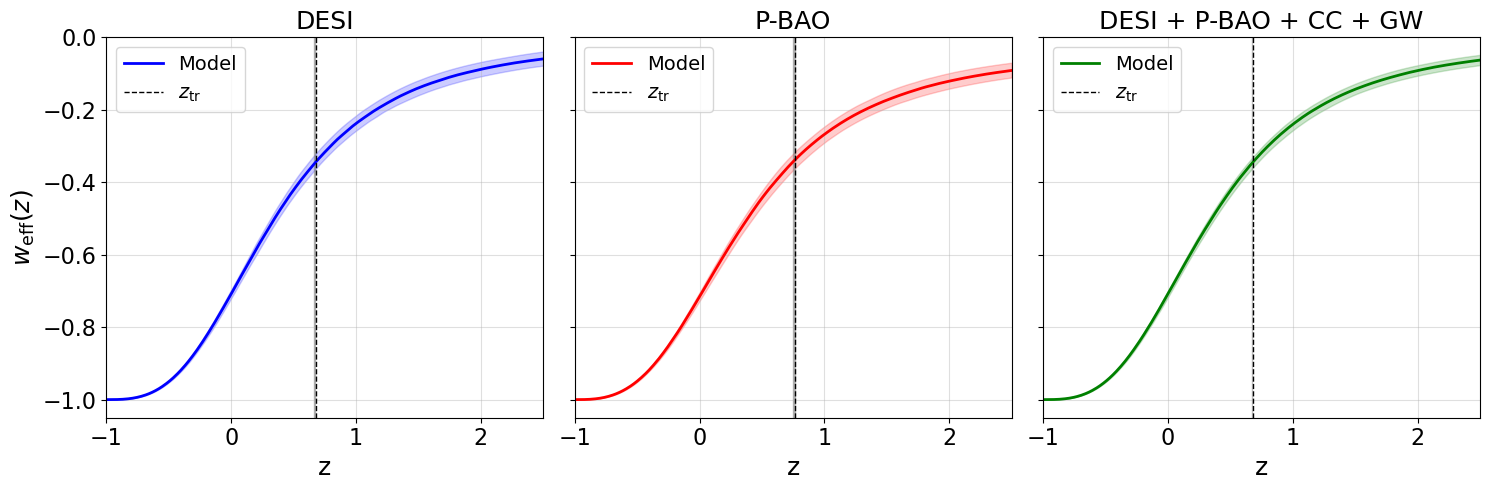}}
\centerline{
\includegraphics[width=0.8\textwidth]{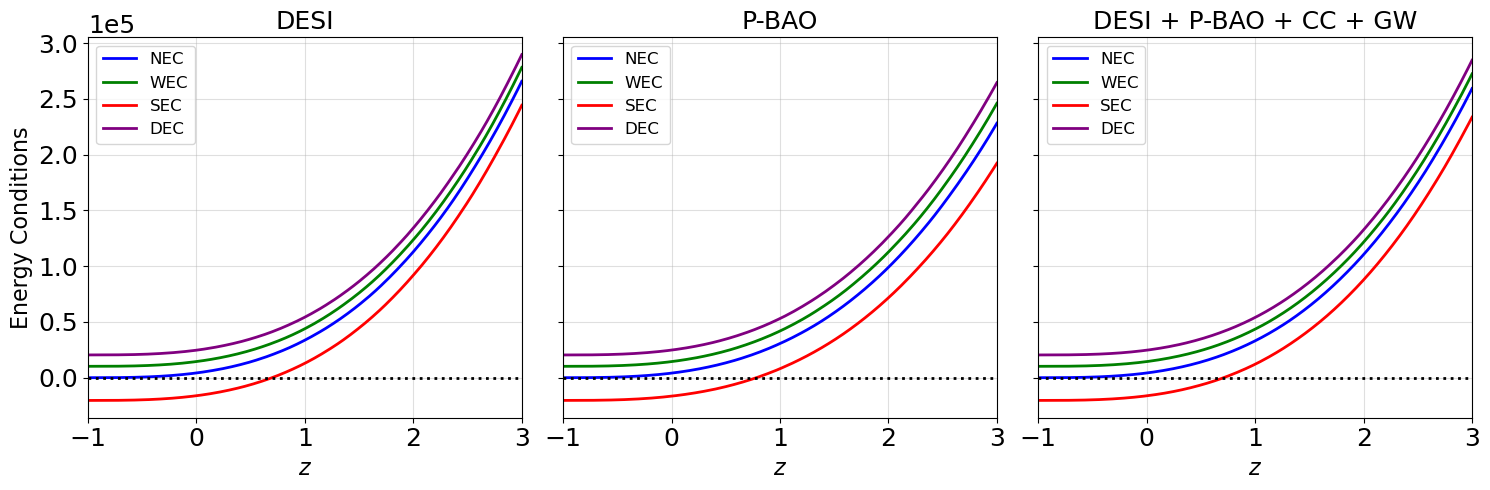}}
\centerline{
\includegraphics[width=0.35\textwidth]{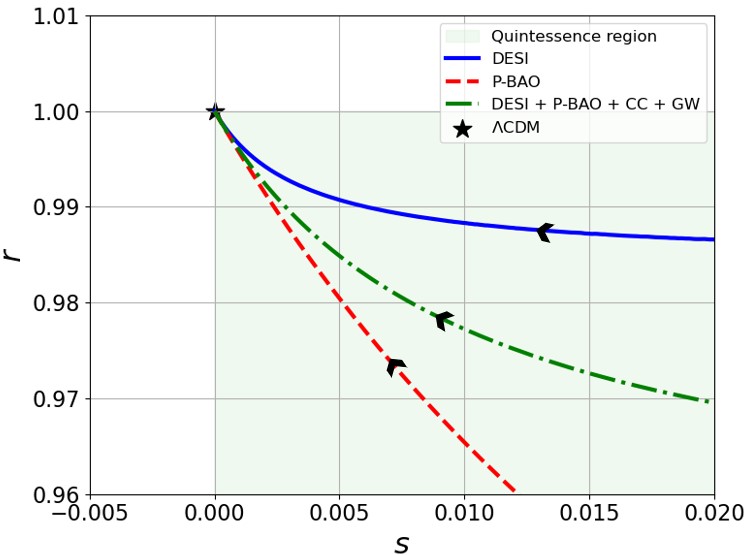}
\includegraphics[width=0.35\textwidth]{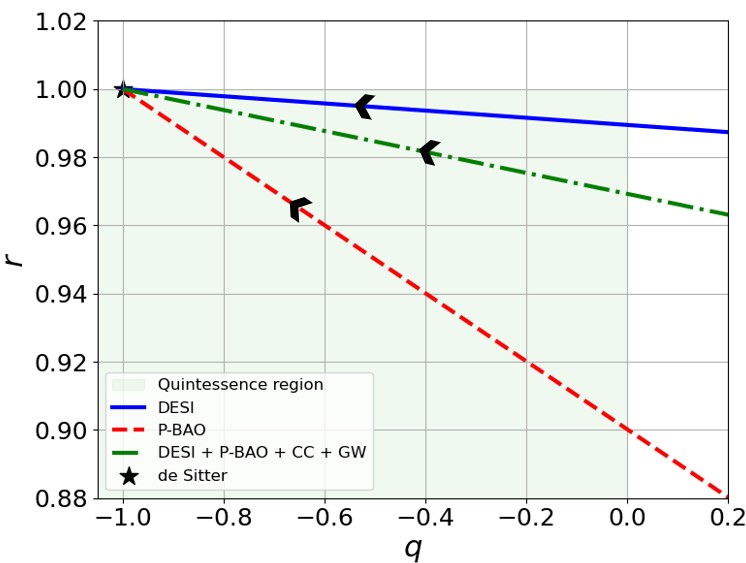}}
\caption{Plot showing evolution of deceleration, effective EoS, and energy conditions with respect to the redshift. Along with the evolution of statefinder trajectories in the $(r,s)$ and $(r,q)$ plane. In top two plots transition redshift in each case is indicated by a dotted line, while the shaded regions represent the allowed range at the 1-$\sigma$ confidence level.}
\label{f3}
\end{figure*}

In this section we put forward the constraints on the given $f(Q,L_m)$ and it's dynamics and validity across both the early and late-time epochs. The results obtained from the MCMC analysis to evaluate the late-time dynamics of the model are summarized in Tables~(\ref{tab1}) and (\ref{tabr1}), along with Figs.~(\ref{f1}) and (\ref{f2}). The model yields a robust and consistent estimate of the Hubble constant, with $H_0 \simeq 69.4~\mathrm{km\,s^{-1}\,Mpc^{-1}}$ across all datasets. This value lies between the Planck and SH0ES measurements \cite{planck2018,verde2019tensions,Riess2021}, potentially alleviating the Hubble tension. The results are also consistent with recent $f(Q)$ and $f(Q,L_m)$ studies \cite{capozziello2022,Sakr,n1}. Furthermore, the deviation of $n$ from unity indicates departures from standard matter evolution, while the stability of $\tilde{\gamma} \sim 3400$ across datasets highlights the robustness of the model. Finally, using the constrained model parameters, we examine the evolution of the deceleration parameter, effective EoS, and statefinder diagnostics. We also analyze the energy conditions to test observational consistency to assess the dynamical viability of the model in the late-time era (Ref. \cite{R1} and Ref. \cite{R2} for detailed explanations).\\
Here, from Fig. (\ref{f3}) and Table (\ref{tabr2}) we observe model successfully describes the late-time accelerated expansion, with $q(0)<0$ and a transition redshift $z_{\rm tr}\sim 0.7$, consistent with other recent observations \cite{pp47,n1, capozziello2022,Sakr}. The effective equation of state remains in the quintessence regime ($-1<\omega_{\rm eff}$) and asymptotically approaches $-1$ without crossing the phantom boundary, ensuring a stable de~Sitter future. At high redshifts, the model recovers the matter-dominated era ($\omega_{\rm eff}\to 0$). The statefinder analysis demonstrates that the model evolves in the quintessence region and gradually approaches the $\Lambda$CDM fixed point $(r,s) = (1,0)$ and the de~Sitter attractor $(r,q) = (1,-1)$ at late times.  The absence of trajectories entering the phantom regime confirms the dynamical consistency and physical viability of the model, while energy conditions such as WEC, NEC, DEC remain satisfied, with SEC violation at late times driving acceleration. Further, from a statistical standpoint, the model consistently achieves lower $\chi^2_{\min}$, AIC, and BIC values than the standard $\Lambda$CDM model across all datasets. The large positive $\Delta$AIC and $\Delta$BIC values, especially for DESI and the combined datasets strongly favor the model, while the comparatively smaller differences for the P-BAO dataset emphasize the crucial role of high-precision data in tightening parameter constraints.\\

\begin{table}[htbp]
\centering
\renewcommand{\arraystretch}{1.25}
\begin{tabular}{lccc}
\toprule
\textbf{Dataset} & \boldmath$H_0$ & \boldmath$n$ & \boldmath$\tilde{\gamma}$ \\
\midrule

DESI & 
$69.42^{+0.47}_{-0.45}$ &
$1.0035^{+0.0184}_{-0.0165}$ &
$3401.93^{+26.52}_{-28.07}$ \\[4pt]

P-BAO & 
$69.36^{+0.47}_{-0.45}$ &
$1.0356^{+0.0223}_{-0.0205}$ &
$3397.13^{+29.07}_{-25.61}$ \\[4pt]

DESI + P-BAO + CC + GW & 
$69.36^{+0.33}_{-0.33}$ &
$1.0105^{+0.0129}_{-0.0122}$ &
$3396.53^{+28.79}_{-25.65}$ \\

\bottomrule
\end{tabular}
\caption{Best-fit parameter values derived from various observational datasets with $1\sigma$ uncertainty.}
\label{tab1}
\end{table}

\begin{table*}[ht!]
\centering
\renewcommand{\arraystretch}{1.25}
\begin{tabular}{llccccc}
\toprule
\textbf{Model} & \textbf{Dataset} & \textbf{$\chi^2_{\min}$} & \textbf{AIC} & \textbf{BIC} & \textbf{$\Delta$AIC} & \textbf{$\Delta$BIC} \\
\midrule

{\textbf{Model}} 
  & DESI               & 15.42 & 19.42 & 21.30 & 25.72 & 25.72 \\
  & P-BAO              & 32.64 & 36.64 & 40.07 & 12.43 & 10.54 \\
  & DESI + P-BAO + CC + GW  & 81.23 & 85.23 & 91.65 & 20.46 & 20.16 \\

\midrule

{\textbf{$\Lambda$CDM}} 
  & DESI               & 41.14 & 45.14 & 47.02 & -- & -- \\
  & P-BAO              & 45.07 & 49.07 & 50.61 & -- & -- \\
  & DESI + P-BAO + CC + GW  & 101.69 & 105.69 & 111.81 & -- & -- \\

\bottomrule
\end{tabular}
\caption{A statistical comparison between THE given $f(Q,L_m)$ model and $\Lambda$CDM framework. We define $\Delta$AIC = $\mathrm{AIC}_{\Lambda\mathrm{CDM}} - \mathrm{AIC}_{\text{model}}$ and $\Delta$BIC = $\mathrm{BIC}_{\Lambda\mathrm{CDM}} - \mathrm{BIC}_{\text{model}}$.}
\label{tabr1}
\end{table*}

\begin{table}[htbp]
\centering
\renewcommand{\arraystretch}{1.25}
\begin{tabular}{lccc}
\toprule
\textbf{Dataset} & \boldmath$q(0)$ & \boldmath$\omega_{\mathrm{eff}}(0)$ & \boldmath$z_{\mathrm{tr}}$ \\
\midrule

DESI 
& $-0.5604^{+0.0017}_{-0.0023}$ 
& $-0.7069^{+0.0011}_{-0.0016}$ 
& $0.6961^{+0.0235}_{-0.0218}$ \\[4pt]

P-BAO 
& $-0.5739^{+0.0031}_{-0.0024}$ 
& $-0.7160^{+0.0021}_{-0.0016}$ 
& $0.7807^{+0.0396}_{-0.0325}$ \\[4pt]

DESI + P-BAO + CC + GW 
& $-0.5635^{+0.0044}_{-0.0033}$ 
& $-0.7090^{+0.0029}_{-0.0022}$ 
& $0.7139^{+0.0194}_{-0.0153}$ \\

\bottomrule
\end{tabular}
\caption{Summary of present-day values of deceleration parameter and the effective EoS parameter from various observational datasets and and the transition redshift $z_{\mathrm{tr}}$ defined by $q(z_{\mathrm{tr}})=0$.}
\label{tabr2}
\end{table}

\begin{figure*}[htb]
\centerline{
\includegraphics[width=0.45\textwidth]{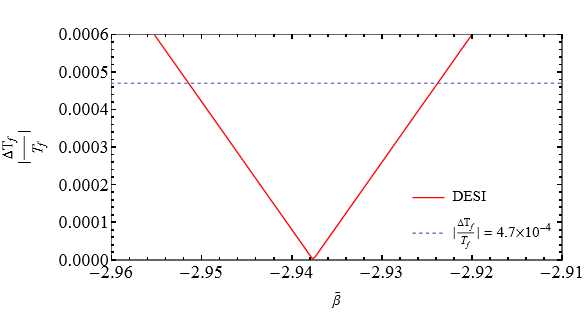}
\includegraphics[width=0.45\textwidth]{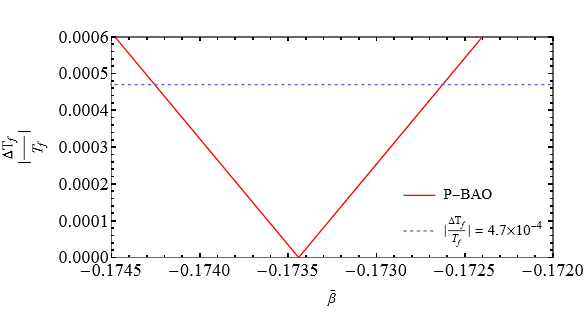}}
\centerline{
\includegraphics[width=0.45\textwidth]{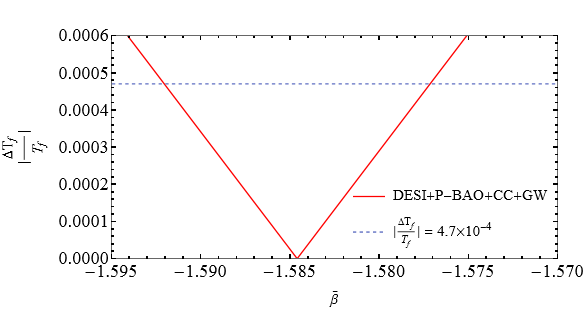}
}
\caption{Plot of $|\frac{\Delta T_{f}}{T_{f}}|$ Vs $\tilde{\beta}$ for constraints on the $\tilde{\beta}$ for which the model remains BBN viable.}
\label{f4t}
\end{figure*}

Now, for analysis of the early time constraints of BBN, we use Eq. (\ref{eqtf}) in Eq. (\ref{Tf_bound}), and evaluate the constraints on the parameter $\tilde{\beta}$ using the best fit values of the model parameters like $n$ and $\tilde{\gamma}$ obtained from the analysis of late-time constraints. The results for which are shown in Fig. (\ref{f4t}) respectively, which clearly shows that for above given constraints on model parameter $n$ and $\tilde{\gamma}$, their exist some range of $\tilde{\beta}$ for which the model is BBN viable.

\section{Conclusion}\label{s7}
In this work, we have carried out a comprehensive analysis of the cosmological dynamics within the framework of $f(Q,L_m)$ gravity by jointly incorporating early- and late-time observational constraints. The model parameters were tightly constrained using DESI DR2, P-BAO, CC, and GW datasets, yielding consistent estimates of the Hubble constant $H_0 \simeq 69.4~\mathrm{km\,s^{-1}\,Mpc^{-1}}$. The deviation of the parameter $n$ from unity signals departures from standard matter evolution, while the stability of $\tilde{\gamma}$ across datasets highlights the robustness of the model. The late-time analysis confirms that the model successfully describes the accelerated expansion of the Universe, with a negative present-day deceleration parameter and a transition redshift $z_{\rm tr} \sim 0.7$, consistent with observational findings. The effective equation of state remains in the quintessence regime ($-1 < \omega_{\rm eff} < 0$) and asymptotically approaches the de~Sitter limit without crossing the phantom boundary, ensuring a stable cosmic evolution. This behavior is further supported by the statefinder diagnostics, which show convergence toward the $\Lambda$CDM fixed point at late times. Moreover, the satisfaction of NEC, WEC, and DEC, along with late-time violation of SEC, reinforces the physical consistency of the model. From a statistical perspective, the model consistently outperforms the standard $\Lambda$CDM scenario, yielding lower values of $\chi^2_{\min}$, AIC, and BIC across all datasets, with significantly large positive $\Delta$AIC and $\Delta$BIC values for DESI and the combined analysis. This highlights the importance of high-precision data in strengthening model selection. In addition, the inclusion of Big Bang Nucleosynthesis constraints provides an independent early-Universe test, restricting the parameter $\tilde{\beta}$ to a viable range consistent with the late-time best-fit values of $n$ and $\tilde{\gamma}$. This demonstrates that the model remains compatible with primordial nucleosynthesis and the early universe physics.
Overall, the $f(Q,L_m)$ gravity framework offers a consistent and flexible account of cosmic evolution, unifying early- and late-time observations, closely mimicking $\Lambda$CDM while permitting controlled deviations and demonstrating enhanced statistical competitiveness. Future investigations including perturbations, large-scale structure, weak lensing, and CMB observations will further assess the robustness of the model across cosmological probes. In particular, analyses of structure growth and redshift-space distortions, along with upcoming surveys such as \textit{Euclid}~\cite{Euclid2018}, \textit{LSST}~\cite{LSST2019}, and \textit{SKA}~\cite{SKA2015}, will also be essential for distinguishing this framework from standard cosmology and providing deeper insights into the geometric origin of cosmic acceleration.
\section{DATA AVAILABILITY}
All data used in this study are cited in the references and were obtained from publicly available sources.

\end{document}